\documentclass[sigconf,natbib=false,pbalance,screen, nonacm]{acmart}

\usepackage{subcaption}
\usepackage[linesnumbered,ruled,vlined]{algorithm2e}
\DontPrintSemicolon

\SetCommentSty{AlgoCommentStyle}
\SetKwComment{tcp}{//~}{}

\newcommand{\repourl}{%
        https://github.com/tdortman/Cuckoo-GPU/
    }

\RequirePackage[
datamodel=acmdatamodel,
style=acmnumeric, 
]{biblatex}
\setcopyright{cc}
\copyrightyear{2026}

\begin{document}

\title{Cuckoo-GPU: Accelerating Cuckoo Filters on Modern GPUs}
 
\author{Tim Dortmann}
\orcid{0009-0008-5497-5811}
\affiliation{
  \institution{Johannes Gutenberg University Mainz}
  \city{Mainz}
  \country{Germany}
}
\email{research@tdortman.com}

\author{Markus Vieth}
\orcid{0000-0002-7281-3364}
\affiliation{
  \institution{Johannes Gutenberg University Mainz}
  \city{Mainz}
  \country{Germany}
}
\email{vieth@uni-mainz.de}

\author{Bertil Schmidt}
\orcid{0000-0003-2597-8331}
\affiliation{
  \institution{Johannes Gutenberg University Mainz}
  \city{Mainz}
  \country{Germany}
}
\email{bertil.schmidt@uni-mainz.de}

\begin{abstract}
Approximate Membership Query (AMQ) structures are essential for high-throughput systems in databases, networking, and bioinformatics. While Bloom filters offer speed, they lack support for deletions. Existing GPU-based dynamic alternatives, such as the Two-Choice Filter (TCF) and GPU Quotient Filter (GQF), enable deletions but incur severe performance penalties. 
We present {\it Cuckoo-GPU}, an open-source, high-performance Cuckoo filter library for GPUs. Instead of prioritizing cache locality, Cuckoo-GPU embraces the inherently random access pattern of Cuckoo hashing to fully saturate global memory bandwidth. Our design features a lock-free architecture built on atomic compare-and-swap operations, paired with a novel breadth-first search-based eviction heuristic that minimizes thread divergence and bounds sequential memory accesses during high-load insertions.
Evaluated on NVIDIA GH200 (HBM3) and RTX PRO 6000 Blackwell (GDDR7) systems, Cuckoo-GPU closes the performance gap between append-only and dynamic AMQ structures. It achieves insertion, query, and deletion throughputs up to 378$\times$ (4.1$\times$), 6$\times$ (34.7$\times$), and 258$\times$ (107$\times$) higher than GQF (TCF) on the same hardware, respectively, and delivers up to a 350$\times$ speedup over the fastest available multi-threaded CPU-based Cuckoo filter implementation. Moreover, its query throughput rivals that of the append-only GPU-based Blocked Bloom filter --- demonstrating that dynamic AMQ structures can be deployed on modern accelerators without sacrificing performance.

\end{abstract}



\maketitle

\section{Introduction}

In the era of big data, the ability to perform set membership queries is a fundamental requirement for applications ranging from content delivery networks and network intrusion systems to database systems and genomic indexing.
Determining whether an element belongs to a set is a frequent and performance-critical operation. The ability to filter non-member elements before performing expensive I/O or compute operations is also often important for maintaining system throughput in high-performance computing (HPC) pipelines.

While exact data structures provide definitive answers, their memory footprint and computational overhead can make them impractical for massive datasets. This has led to the widespread adoption of probabilistic data structures (filters), which trade a small, controllable {\it false positive rate} (FPR) $\epsilon$ for significant gains in space and time efficiency: 
An {\it approximate membership query} (AMQ) for an item $x$ always returns true if $x \in S$ and returns false with probability $1-\epsilon$ if $x \notin S$.

For decades, the Bloom filter \cite{bloom} has been the standard probabilistic data structure for AMQ despite its inability to delete elements, making it unsuitable for dynamic datasets. This has led to the design of a variety of alternative filters with examples including Cuckoo filters \cite{cuckoo-filter}, Quotient filters \cite{quotient-filter}, Two-Choice filters \cite{gpu-filters}, Morton filters \cite{breslow2018morton}, XOR filters \cite{graf2020xor}, and Ribbon filters \cite{dillinger2021ribbon}.  
While some of these variations do support deletion, they can incur prohibitive space overheads that often negate their practical viability.
Due to this the Cuckoo filter has emerged as a powerful alternative, offering native support for deletions and often features superior space efficiency at low FPR.

GPUs are a popular accelerator architecture across a variety of data-intensive applications. 
Even though they provide immense compute power through massive parallelism, data movement frequently becomes a major bottleneck. 
Thus, utilising filters directly on GPUs can be highly attractive to reduce data movements in processing pipelines.  

While optimisations for Cuckoo filters have been well-studied on CPUs \cite{lang2019performance, schmidt2021four}, porting them to GPUs presents significant challenges. 
The insertion algorithm relies heavily on random memory accesses.
This is only exacerbated by the sequential nature of its eviction chains, where displacing one item can trigger a cascade of evictions. 
These chains can grow quite large, fundamentally conflicting with the GPU's desire for coalesced memory access and minimal thread divergence.

Recent attempts to adapt dynamic AMQ data structures for GPUs, such as the Two-Choice filter (TCF) \cite{gpu-filters} and the GPU Quotient filter (GQF) \cite{gpu-qf,gpu-filters} have tried to mitigate these issues by prioritising memory locality over all else. 
The GQF, for example, relies on element shifting within contiguous blocks to maintain sorted order, while the TCF utilises complex cooperative group scheduling.
Although these strategies successfully improve cache hit rates, they also introduce significant overheads.
As a result, these filters fail to saturate the massive global memory bandwidth available on modern High-Bandwidth Memory (HBM) architectures.

In this paper, we present the design and evaluation of {\it Cuckoo-GPU} --- the first high-performance GPU-based Cuckoo filter library. 
Rather than over-optimising for memory locality at the expense of algorithmic overhead, our approach embraces the random-access nature of Cuckoo hashing, relying on lock-free concurrency to fully leverage the massive memory bandwidth of modern GPUs. 
We make the following contributions in this work: 

\begin{itemize}
    \item {\bf High-Performance CUDA Library: } We present the design and implementation of a parallel Cuckoo filter supporting insertion, query, and deletion based on atomic {\it compare-and-swap} (CAS). \footnote{The source code is available as an open-source header-only library at: \url{\repourl}}
    
    \item {\bf Advanced Optimisation Techniques:} We explore and evaluate several optimisation strategies to optimize occupancy and memory bandwidth. 
    These include a novel Breadth-first search (BFS) eviction strategy designed to reduce the number of random memory accesses at high load factors, as well as a flexible offset-based bucket placement policy to eliminate memory over-provisioning.
    
    \item {\bf Comprehensive Evaluation:} Through a rigorous evaluation on an NVIDIA GH200 (HBM3) and RTX PRO 6000 Blackwell (GDDR7), we demonstrate that our approach scales exceptionally well with global memory bandwidth, whereas state-of-the-art competitors stagnate due to internal bottlenecks. 
    \textit{Cuckoo-GPU} achieves insertion, query, and deletion throughputs up to 378$\times$ (4.1$\times$), 6$\times$ (34.7$\times$), and 258$\times$ (107$\times$) higher than GQF (TCF) when executed on the same hardware, respectively, and delivers up to a 350$\times$ speedup over the fastest available multi-threaded CPU-based Cuckoo filter implementation.
    Furthermore, it rivals the query throughput of the append-only GPU-based Blocked Bloom filter while far surpassing all other tested dynamic filters.
\end{itemize}

The rest of the paper is organized as follows.
After providing background in Section \ref{sec:background} and reviewing related work in Section \ref{sec:related}, we present the design of Cuckoo-GPU in Section \ref{sec:design}. 
We perform an extensive experimental evaluation against a set of state-of-the-art baselines on both CPUs and GPUs in Section \ref{sec:eval}. 
Finally, we conclude in Section \ref{sec:conclusion}.

\section{Background}
\label{sec:background}

\subsection{Cuckoo filter}
\label{sec:background:cuckoo-filter}

\begin{figure}
	\centering
	\includegraphics[width=.85\linewidth]{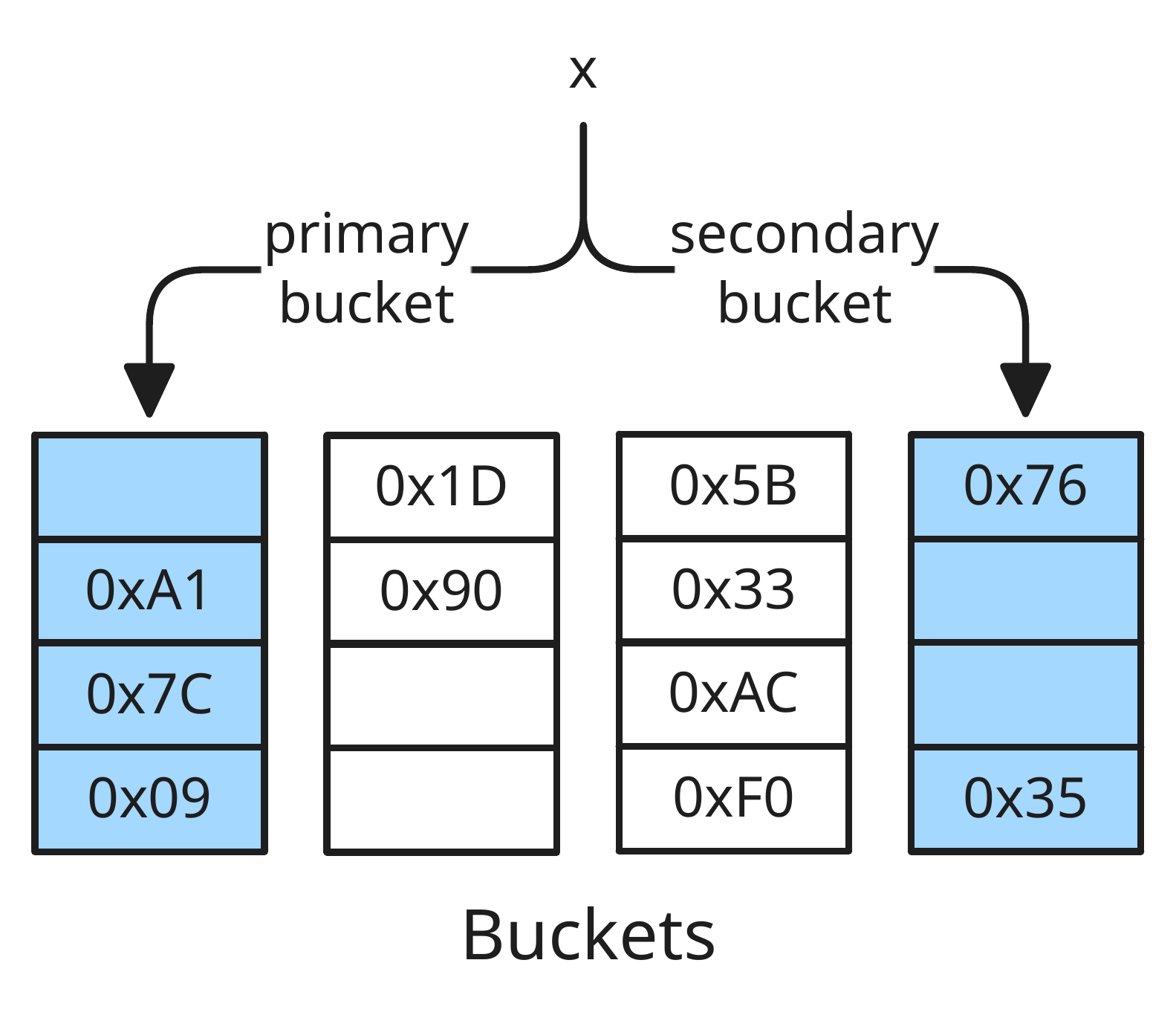}
	\caption{Cuckoo filter with 8-bit fingerprints ($f=8$) per bucket and each bucket containing up to $b = 4$ fingerprints.}
    \Description{Four buckets shown: first bucket contains 0xA1, 0x7C, 0x09; second bucket contains 0x5B, 0x33, 0xAC, 0xF0; third bucket contains 0x1D, 0x90; fourth bucket contains 0x76, 0x35. Element 'x' maps to the first bucket as primary and the fourth bucket as secondary bucket via hash functions.}
	\label{fig:cuckoo}
\end{figure}

The Cuckoo filter \cite{cuckoo-filter} is a probabilistic data structure that improves upon the Bloom filter by supporting item deletion and offering higher space efficiency at low FPR. Instead of setting bits in an array, the Cuckoo filter stores a small, fixed-size fingerprint $fp$ of size $f$ bits for each item. 
The filter consists of an array of buckets, where each bucket can hold up to a fixed number of $b$ fingerprints.
It can be seen as a compact variant of a Cuckoo hash table \cite{cuckoo-hashing}, where an element can only be inserted in one of two possible buckets. Fig. \ref{fig:cuckoo} illustrates an example with $b=4$ and $f=8$. By allowing items to be relocated during insertion, Cuckoo filters can achieve very high load factors. 

To support element relocation without storing the original keys, Cuckoo filters utilise \textit{partial-key cuckoo hashing}.
Let $\mathcal{H}$ represent a uniform, non-cryptographic hash function.
For an item $x$, the tag  $fp$ and the two candidate bucket indices $i_1$ and $i_2$ are calculated as:

\begin{align}
    fp &= \text{fingerprint(}\mathcal{H}(x)) \\
    i_1 &= \mathcal{H}(x) \\
    i_2 &= i_1 \oplus \mathcal{H}(fp)
\end{align}

The XOR operation ($\oplus$) ensures that $i_1$ can also be computed from $i_2$ and the fingerprint the same way, enabling the relocation of items during insertion. This allows an evicted fingerprint to calculate its alternate location from its current position and its own fingerprint alone, an important feature since the original item is not stored.

\textbf{Insertion and Eviction.} To insert an item, the algorithm checks if either bucket $i_1$ or $i_2$ has an empty slot. 
If both are full, a random fingerprint is evicted from one of the buckets and swapped with the new item. 
The displaced fingerprint is then moved to its alternate location. This process may trigger a cascading chain of evictions until an empty slot is found or a maximum depth is reached.

\textbf{Query and Deletion.} A query checks both candidate buckets for the corresponding fingerprint. 
Deletion is performed by removing a matching fingerprint from either bucket when found. 
Unlike Bloom filters, this can be done safely without affecting other items, provided that fingerprint collisions are handled or accepted as a cause of false deletions with a small probability. 

\textbf{False Positive Rate.} The FPR $\epsilon$ of a Cuckoo filter can be approximated by \cite{lang2019performance}:

\begin{equation}
\epsilon \approx 1 - \left(1 - 2^{-f} \right)^{2b\alpha},
\label{eq:cuckoo-fpr}
\end{equation}

whereby $\alpha$ denotes the load factor of the hash table. It is directly tied to the fingerprint size $f$ and the bucket size $b$; i.e., $\epsilon$ improves as $b$ decreases and $f$ increases.

\subsection{Modern GPU Architectures}

 Processing on the GH200 (RTX PRO 6000 Blackwell) GPU is distributed across 132 (188) streaming multiprocessors (SMs) containing 4 vector units of 32 cores. Typically, one unit of work is assigned to one thread. When running a multi-threaded task, the GPU distributes small batches of threads (thread blocks) across SMs, which then schedule even smaller batches of 32 threads (warps) onto the 32-core vector units. Each SM has its own L1 cache (128 KB), which can be partially reconfigured to serve as fast user-programmable shared memory, allowing for fast data movement and storage within a thread block. Similarly, each SM contains 256 KB worth of 32-bit registers, which can be cross-referenced by threads within a warp to exchange data via warp shuffling. Additionally, the GPU features a unified L2 cache (50 MB (128 MB) for GH200 (RTX PRO 6000)). When multiple neighbouring threads within a warp load neighbouring entries at the same time, the GPU performs a single, larger memory access instead, usually referred to as \textit{access coalescing}.

 To ensure scalability, direct communication across SMs is restricted to communication though the GPU global main memory, which is based on GDDR or high-bandwidth on-chip memory (HBM) on modern GPUs. 
 While the former typically achieve just above one terabyte per second of throughput, e.g., 1.8~TB/s on our utilized RTX PRO 6000, HBM is capable of sustaining several terabytes per second of throughput in the optimal case, e.g., 3.4~TB/s on our utilized GH200. This bandwidth, however, comes at the cost of relatively high access latency. 
The minimum access granularity is 32B, referred to as a sector, with four sectors composing a 128B cache line. 
Accesses from threads on the same SM that target the same cache line or even sector (and thus the same L2 slice) can be merged into a single L2 request through temporal coalescing. Consequently, coalesced memory access patterns are critical for efficient use of GPU DRAM, as they minimize the number of high-latency memory transactions. Importantly, atomic updates benefit from the same coalescing mechanism as well, which is essential for supporting concurrent, lock-free insertions into the Cuckoo filter.
  
\section{Related Work}
\label{sec:related}

\textbf{Bloom Filter Variants.} 
The standard Bloom filter \cite{bloom} and its cache-friendly \textit{Blocked Bloom filter} (BBF) \cite{blocked-bloom} variant provide fast AMQs but lack deletion support. 
Recently, Jünger et al. \cite{jünger2025optimizingbloomfiltersmodern} proposed highly optimised GPU Bloom filters that fully saturate memory bandwidth via vectorization and thread cooperation. 
As their source code is not yet public, we utilize the standard \textit{cuCollections} \cite{cuCollections} BBF as our high-performance static baseline.

\textbf{Dynamic GPU Filters.}
To support deletions on GPUs, recent data structures prioritise memory locality but incur severe architectural trade-offs. 
The \textit{Two-Choice filter} (TCF) \cite{gpu-filters} adapts the power-of-two-choices paradigm \cite{potc}, eliminating eviction chains by relegating overflows to a secondary stash. 
However, its reliance on CUDA Cooperative Groups to cooperatively load and sort blocks in shared memory introduces massive compute and intra-warp synchronisation overheads, preventing it from scaling on high-bandwidth architectures like HBM3.

Alternatively, the \textit{GPU Counting Quotient filter} (GQF) \cite{gpu-filters} improves upon earlier, highly constrained GPU implementations \cite{gpu-qf} by utilising an "even-odd" lock-free strategy to compactly store fingerprints via Robin Hood hashing \cite{robin-hood-hashing}. 
However, maintaining sorted contiguous runs requires shifting elements during updates. This creates strict serial dependencies between threads, making the GQF fundamentally latency-bound.

\textbf{Cuckoo Filters and Hash Tables.} 
While Cuckoo filters have been heavily optimised for CPUs \cite{lang2019performance, schmidt2021four, schmitz2026smaller}, GPU adaptations remain scarce and highly application-specific (e.g., deep packet inspection \cite{ho2018parallel}). 
Instead of probabilistic filters, AMQs can also be answered deterministically using exact data structures like the highly optimised GPU Bucketed Cuckoo Hash Table (BCHT) \cite{awad2023analyzing}. 
However, as we will demonstrate, storing full keys rather than compact fingerprints yields severe memory and throughput penalties.

\textbf{Applications.} 
High-throughput dynamic AMQs are critical for domains spanning network security \cite{grashofer2018towards}, distributed databases \cite{yao2023mdcf}, and bioinformatics \cite{gaia2019ngsreadstreatment, zentgraf2026cleanifier}. 
To demonstrate the practical utility of \textit{Cuckoo-GPU} beyond synthetic benchmarks, we include genomic $k$-mer indexing as a real-world case study in Section \ref{sec:eval}.

\section{Design of Cuckoo-GPU}
\label{sec:design}

In this section, we discuss the design principles necessary to build a high-performance Cuckoo filter on the GPU and provide an overview of our parallelization and implementation, \textit{Cuckoo-GPU}. 
We analyse how the filter's architecture addresses the unique constraints of the GPU memory subsystem.

\subsection{Design Principles}

Adapting dynamic data structures like Cuckoo filters to GPUs requires navigating a few important design principles:

\begin{enumerate}
    \item \textbf{Low Thread Divergence:} Threads within a warp should follow the same execution path. 
    Standard Cuckoo insertion creates high divergence due to random eviction chains, where some threads succeed immediately while others enter long loops.
    \item \textbf{High Memory Coherence:} Threads within a warp should access contiguous memory addresses to enable coalescing.
    \item \textbf{High Degree of Parallelism:} To saturate memory bandwidth and hide latency, the algorithms must expose massive parallelism.
    \item \textbf{Atomic Operations:} Effective synchronisation is crucial. Non-atomic algorithms require locks, which introduce massive overhead.
\end{enumerate}

\subsection{Data Layout}
The filter's primary data storage is a single, contiguous array of fixed-size buckets allocated in GPU global memory.
To maximise memory bandwidth and avoid misaligned access, the internal layout is carefully structured. 
Each bucket is composed of an array of 64-bit unsigned integer words, with (configurable) fingerprints (tags) tightly packed into these 64-bit words.
For example, a 64-bit word can hold eight 8-bit fingerprints or four 16-bit fingerprints.

\begin{figure}[htbp]
  \centering
  \includegraphics[width=\linewidth]{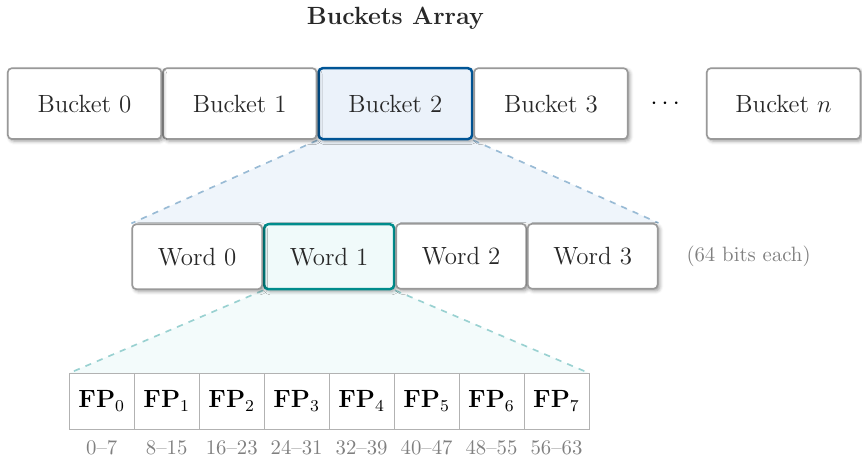}
  \caption{Memory layout of the GPU-based Cuckoo filter.
  Data is hierarchically structured from buckets to words, to tightly packed fingerprints.}
  \Description{GPU Cuckoo filter memory layout: buckets in the array map to 4 words (64 bits each), with 8 fingerprints (FP₀–FP₇) tightly packed per word, indexed by bit ranges 0–7 to 56–63.}
  \label{fig:memory-layout}
\end{figure}

While this packed layout necessitates the use of bitwise shift and mask operations to extract individual fingerprints, the additional computational cost of these operations was found to be negligible compared to the latency of memory access, making this a highly beneficial trade-off.
Figure \ref{fig:memory-layout} shows our hierarchical memory layout.

\subsection{Insertion}

The parallel insertion algorithm is designed to handle a large batch of items in parallel, with each CUDA thread being responsible for inserting a single item.
The process for each thread is shown in Algorithm \ref{alg:parallel-insertion} and can be described as follows:

\begin{enumerate}
    \item \textbf{Hashing and Key Generation}: Each item is first hashed into a 64-bit value using the xxHash64 algorithm, chosen for its high performance and excellent statistical properties.
    This hash is split: the upper 32 bits derive the fingerprint, and the lower 32 bits determine the primary bucket index.
    Distinct hash parts are used to avoid fingerprint clustering.
    The alternate bucket index is then calculated using the partial-key Cuckoo hashing scheme (see Section \ref{sec:background:cuckoo-filter}).

    \item \textbf{Direct Insertion Attempt}: The thread checks the two candidate buckets.
    To distribute items evenly and reduce contention on the first slots of a bucket, the thread does not start scanning at index 0.
    Instead, it uses the item's fingerprint to calculate a pseudo-random starting word index.
    It then iterates through the bucket's words, wrapping around to the beginning.
    For each word, it utilises a bitwise SWAR (SIMD Within A Register) algorithm \cite{bit-twiddling-hacks} to generate a mask of empty slots.
    If a slot is found, an atomic {\it Compare-And-Swap} (CAS) attempts to insert the fingerprint.

    \item \textbf{Eviction Process}: If both candidate buckets are full, the thread initiates the eviction process.
    It randomly selects one bucket and a random occupied slot within it.
    The new fingerprint is then atomically swapped into the chosen slot.
    The evicted fingerprint becomes the new item to insert, and the thread calculates its alternate bucket to continue the process.

    \item \textbf{Termination}: The eviction loop continues until an empty slot is found or a limit on the number of evictions is reached, triggering an insertion failure.
\end{enumerate}

Finally, to maintain an accurate occupancy count without bottlenecking on a single global atomic variable, we employ a hierarchical reduction. 
Successful insertions are tallied at warp level using fast shuffle instructions, aggregated in shared memory at the block level, and ultimately committed via a single atomic addition to global memory per block.

\begin{algorithm}
  \caption{Parallel Insertion}
  \label{alg:parallel-insertion}
  \small
  \SetKwProg{Fn}{Function}{}{}
  \Fn{Insert(key)}{
    \texttt{h} $\gets$ \texttt{Hash}(\texttt{key})\;
    \texttt{fp} $\gets$ \texttt{Fingerprint}(\texttt{h})\;
    $i_1$ $\gets$ \texttt{PrimaryIndex}(\texttt{h})\;
    $i_2$ $\gets$ \texttt{AlternateIndex}($i_1$, \texttt{fp})\;

    \tcp*[l]{Phase 1: Try direct insertion}
    \If{TryInsert($i_1$, fp) \textbf{or} TryInsert($i_2$, fp)}{
      \KwRet \texttt{Success}\;
    }

    \tcp*[l]{Phase 2: Eviction chain}
    \texttt{b} $\gets$ randomly pick $i_1$ or $i_2$\;
    \texttt{tag} $\gets$ \texttt{fp}\;

    \For{$n = 1$ \KwTo maxEvictions}{
      \texttt{s} $\gets$ random slot index in bucket \texttt{b}\;
      \texttt{wordIdx} $\gets$ \texttt{s} / \texttt{tagsPerWord}\;
      \texttt{slotIdx} $\gets$ \texttt{s} $\bmod$ \texttt{tagsPerWord}\;
      \texttt{word} $\gets$ \texttt{buckets}[\texttt{b}][\texttt{wordIdx}]\;

      \tcp*[l]{Atomically swap current tag with existing tag}
      \texttt{casOk} $\gets$ \texttt{false}\;
      \While{casOk is false}{
        \texttt{evicted} $\gets$ \texttt{ExtractTag}(\texttt{word}, \texttt{slotIdx})\;
        \texttt{desired} $\gets$ \texttt{ReplaceTag}(\texttt{word}, \texttt{slotIdx}, \texttt{tag})\;
        \texttt{casOk} $\gets$ \texttt{AtomicCAS}(\&\texttt{buckets}[\texttt{b}][\texttt{wordIdx}], \texttt{word}, \texttt{desired})\;
      }

      \tcp*[l]{Try to insert evicted tag into its alternate bucket}
      \texttt{tag} $\gets$ \texttt{evicted}\;
      \texttt{b} $\gets$ \texttt{AlternateIndex}(\texttt{b}, \texttt{tag})\;

      \If{TryInsert(buckets[b], tag)}{
        \KwRet \texttt{Success}\;
      }
    }

    \tcp*[l]{Table too full, caller will have to rebuild}
    \KwRet \texttt{Failure}\;
  }
  \BlankLine
  \Fn{TryInsert(bucket, tag)}{
    \texttt{start} $\gets$ (\texttt{tag} $\bmod$ \texttt{bucketSize}) / \texttt{tagsPerWord}\;

    \For{$i = 0$ \KwTo wordsPerBucket $- 1$}{
      \texttt{idx} $\gets$ (\texttt{start} $+$ \texttt{i}) $\bmod$ \texttt{wordsPerBucket}\;
      \texttt{word} $\gets$ \texttt{bucket}[\texttt{idx}]\;
      \texttt{mask} $\gets$ \texttt{ZeroMask}(\texttt{word})\;

      \While{mask is not 0}{
        \texttt{slot} $\gets$ \texttt{FindFirstSet}(\texttt{mask})\;
        \texttt{desired} $\gets$ \texttt{ReplaceTag}(\texttt{word}, \texttt{slot}, \texttt{tag})\;
        \If{AtomicCAS(\&bucket[idx], word, desired)}{
          \KwRet \texttt{Success}\;
        }
        \tcp*[l]{Reload on CAS failure}
        \texttt{word} $\gets$ \texttt{bucket}[\texttt{idx}]\;
        \texttt{mask} $\gets$ \texttt{ZeroMask}(\texttt{word})\;
      }
    }
    \KwRet \texttt{Failure}\;
  }
\end{algorithm}

\subsection{Query}
\label{sec:design:query}

The parallel query algorithm is a read-only operation optimised for memory access.
Each CUDA thread calculates the fingerprint and two bucket indices as shown in Algorithm \ref{alg:parallel-query}.
Similar to insertion, the thread determines a random starting word based on the fingerprint to avoid checking the same memory locations first for every query.
Our key optimisation is vectorised, non-atomic memory loads combined with SWAR comparisons.
Depending on the hardware and instruction path, a single load can return multiple 64-bit words (for example, two words for 128-bit loads or four words for 256-bit loads).
The kernel therefore models loading generically: \texttt{LoadWords()} returns a sequence of words from an aligned offset, and the algorithm iterates over that sequence through an unrolled loop.
Each returned word is compared against the broadcast fingerprint using constant-time arithmetic, eliminating branching loops.

Modern architectures (e.g., NVIDIA Blackwell with Compute Capability 10.0 or later), can use dedicated 256-bit load instructions (e.g., \texttt{ld.global.nc.v4.u64}).\footnote{NVIDIA PTX ISA Documentation: https://docs.nvidia.com/cuda/parallel-thread-execution/index.html\#data-movement-and-conversion-instructions-ld}
This instruction fetches four \texttt{uint64\_t} values in one operation, bypassing the L1 cache to access L2 cache or global memory directly.
In cache-resident scenarios, this can reduce instruction count and pressure on the instruction pipeline, resulting in a "free" performance boost.

Because the vectorised memory loads are non-atomic and non-coherent, the query kernel cannot safely execute concurrently with insertions or deletions without risking torn reads.

\begin{algorithm}
  \caption{Parallel Query}
  \label{alg:parallel-query}
  \small
  \SetKwProg{Fn}{Function}{}{}
  \Fn{Contains(key)}{
    \texttt{h} $\gets$ \texttt{Hash}(\texttt{key})\;
    \texttt{fp} $\gets$ \texttt{Fingerprint}(\texttt{h})\;
    $i_1$ $\gets$ \texttt{PrimaryIndex}(\texttt{h})\;
    $i_2$ $\gets$ \texttt{AlternateIndex}($i_1$, \texttt{fp})\;

    \tcp*[l]{Check both buckets (read-only)}
    \KwRet \texttt{Find}($i_1$, \texttt{fp}) \textbf{or} \texttt{Find}($i_2$, \texttt{fp})\;
  }
  \BlankLine
  \Fn{Find(bucket, tag)}{
    \texttt{pattern} $\gets$ \texttt{BroadcastTag}(\texttt{tag})\;

    \tcp*[l]{Random start index aligned to current load width}
    \texttt{loadWidthWords} $\gets$ \texttt{WordLoadWidth}()\;
    \texttt{start} $\gets$ (\texttt{tag} $\bmod$ \texttt{bucketSize}) / \texttt{tagsPerWord}\;
    \texttt{start} $\gets$ \texttt{AlignDown}(\texttt{start}, \texttt{loadWidthWords})\;

    \texttt{i} $\gets 0$\;
    \While{$i <$ wordsPerBucket}{
      \texttt{idx} $\gets$ (\texttt{start} $+$ \texttt{i}) $\bmod$ \texttt{wordsPerBucket}\;
      \texttt{words} $\gets$ \texttt{LoadWords}(\texttt{bucket}, \texttt{idx})\;

      \tcp*[l]{Use SWAR to check all loaded words}
      \For{$w$ \textbf{in} words}{
        \If{HasZeroSegment($w \oplus$ pattern)}{
          \KwRet \texttt{Success}\;
        }
      }

      \texttt{i} $\gets$ \texttt{i} $+$ \texttt{loadWidthWords}\;
    }
    \KwRet \texttt{Failure}\;
  }
\end{algorithm}

\subsection{Deletion}

The parallel deletion algorithm leverages SWAR to locate and remove items efficiently.
Like the other operations, it iterates through the bucket in 64-bit words, starting at a pseudo-random offset derived from the fingerprint as shown in Algorithm \ref{alg:parallel-deletion}.

Each thread broadcasts the target tag and uses SWAR to find matches.
If a match is found, it attempts an atomic CAS to set the specific slot to \texttt{EMPTY} (zero).
If the CAS fails (due to concurrent modification), the thread reloads and retries.
This ensures thread safety without locking.
The operation continues until the item is removed, or the entire bucket has been scanned.

\begin{algorithm}
  \caption{Parallel Deletion}
  \label{alg:parallel-deletion}
  \small
  \SetKwProg{Fn}{Function}{}{}
  \Fn{Remove(key)}{
    \texttt{h} $\gets$ \texttt{Hash}(\texttt{key})\;
    \texttt{fp} $\gets$ \texttt{Fingerprint}(\texttt{h})\;
    $i_1$ $\gets$ \texttt{PrimaryIndex}(\texttt{h})\;
    $i_2$ $\gets$ \texttt{AlternateIndex}($i_1$, \texttt{fp})\;

    \tcp*[l]{Attempt to remove from either valid location}
    \KwRet \texttt{TryRemove}($i_1$, \texttt{fp}) \textbf{or} \texttt{TryRemove}($i_2$, \texttt{fp})\;
  }
  \BlankLine
  \Fn{TryRemove(bucket, targetTag)}{
    \texttt{start} $\gets$ (\texttt{targetTag} $\bmod$ \texttt{bucketSize}) / \texttt{tagsPerWord}\;
    \texttt{pattern} $\gets$ \texttt{BroadcastTag}(\texttt{targetTag})\;

    \For{$i = 0$ \KwTo wordsPerBucket $- 1$}{
      \texttt{idx} $\gets$ (\texttt{start} $+$ \texttt{i}) $\bmod$ \texttt{wordsPerBucket}\;
      \texttt{word} $\gets$ \texttt{bucket}[\texttt{idx}]\;
      \texttt{mask} $\gets$ \texttt{ZeroMask}(\texttt{word} $\oplus$ \texttt{pattern})\;

      \While{mask is not 0}{
        \texttt{slot} $\gets$ \texttt{FindFirstSet}(\texttt{mask})\;
        \texttt{desired} $\gets$ \texttt{ReplaceTag}(\texttt{word}, \texttt{slot}, \texttt{EMPTY})\;

        \If{AtomicCAS(\&bucket[idx], word, desired)}{
          \KwRet \texttt{Success}\;
        }

        \tcp*[l]{Reload on CAS failure}
        \texttt{word} $\gets$ \texttt{bucket}[\texttt{idx}]\;
        \texttt{mask} $\gets$ \texttt{ZeroMask}(\texttt{word} $\oplus$ \texttt{pattern})\;
      }
    }
    \KwRet \texttt{Failure}\;
  }
\end{algorithm}

\subsection{Optimization Strategies}

\subsubsection{\textbf{Alternate Eviction Strategy}}
\label{sec:impl:eviction-strategies}
The standard eviction process uses a greedy, depth-first-search (DFS) approach, where a thread immediately follows the eviction chain of a single evicted item.
To reduce the average length of these chains and mitigate thread divergence, we implemented an alternate strategy based on a breadth-first-search (BFS) heuristic. 
When an eviction is necessary, instead of picking one random item to evict, the thread inspects up to half the items in the current bucket. 
For each candidate, it checks whether its respective alternate bucket contains an empty slot. 

If such an eviction path is found, the algorithm executes a two-step lock-free relocation. 
First, it inserts the candidate fingerprint into the discovered empty slot in its alternate bucket. 
Next, it attempts to atomically replace the candidate in the original bucket with the new fingerprint using a CAS operation. 
If this CAS fails, indicating that the original slot was modified by another thread, the algorithm must remove the candidate fingerprint it just inserted to avoid creating duplicates.

If all inspected candidates lead to full alternate buckets, the algorithm evicts the last candidate that was checked. 
The thread then carries this evicted fingerprint to its alternate bucket and restarts the BFS process from there.
Note that long eviction chains become more common with higher load factors (e.g., 90\% or more). Thus, we will evaluate the impact of the BFS eviction strategy in comparison to DFS depending on the load factor in Section \ref{sec:eval:eviction-policies}.

\subsubsection{\textbf{Flexible Bucket Placement Policies}}
\label{sec:impl:bucket-policies}

For the standard partial-key Cuckoo hashing scheme to map validly onto the filter's buckets, their number must strictly be a power of two \cite{cuckoo-false-negatives}.
This constraint introduces a significant memory footprint issue: if a dataset requires slightly more capacity than $2^n$, the filter must be sized to $2^{n+1}$, resulting in significantly higher memory usage.

To mitigate this over-provisioning, an alternative offset-based bucket placement policy was implemented to decouple the capacity from powers of two. 
Derived from the work of Schmitz et al. \cite{schmitz2026smaller}, we use an asymmetric offset combined with a "choice bit" to indicate the current location of the fingerprint.

\begin{itemize}
  \item If the choice bit is 0, the item is in its primary bucket $i_1$.
    The alternate bucket is calculated as:
    $$ i_2 = (i_1 + \text{offset}(fp)) \bmod m $$
    
  \item If the choice bit is 1, the item is in its alternate bucket $i_2$.
    The primary bucket is calculated as:
    $$ i_1 = (i_2 - \text{offset}(fp)) \bmod m $$
\end{itemize}

When an item is moved between buckets during an eviction, its choice bit is flipped.
This approach supports any number of buckets $m$, offering maximal space efficiency.
The trade-off is a reduction in the effective fingerprint size by one bit (increasing the FPR slightly) and the need to update the choice bit during evictions. We evaluate this trade-off in Section \ref{sec:eval:bucket-policies}.

\subsubsection{\textbf{Sorted Insertion}}
\label{sec:impl:sorted-insertion}

To strictly satisfy the design principle of high memory coherence, we also explored a pre-sorted insertion variant.
By leveraging CUB's high-performance radix sort to order the input batch by their primary bucket index prior to kernel execution, threads within a warp naturally access contiguous memory regions --- at least until eviction chains begin.
While this technique effectively mitigates random memory access penalties, we found that the computational overhead of the sorting fails to amortise on modern GPUs.
Because high-bandwidth memory naturally absorbs much of the uncoalesced access penalty, \textit{Cuckoo-GPU} defaults to the standard, unsorted approach to maximise overall throughput on state-of-the-art accelerators.

\subsection{Library Design}

To facilitate seamless integration into existing GPU-accelerated applications, \textit{Cuckoo-GPU} is implemented as a robust, header-only CUDA library. 
Developers can simply include the library in their projects without complex build system modifications. 
The public API is designed for flexibility and ease of use, providing two distinct interfaces: a traditional C-style API that operates on raw device pointers and batch lengths, and a set of overloaded methods that accept \texttt{thrust::device\_vector} objects. 

To maximise performance, we strictly avoid runtime overhead wherever possible. 
All critical parameters are exposed as compile-time constants through a single template configuration structure. 
This allows the compiler to generate highly specialised code and statically unroll loops.

\section{Evaluation}
\label{sec:eval}
\subsection{Experimental Setup}
The performance evaluation was conducted on three distinct hardware configurations.

\begin{itemize}
    \item \textbf{System A}: An AMD EPYC 7713P (64 cores) paired with an RTX PRO 6000 Blackwell GPU featuring 96 GB of GDDR7 memory (1.8 TB/s) running AlmaLinux 10.1 and CUDA 12.9.
    
    \item \textbf{System B}: A GH200 Grace Hopper system with 72 ARM Neoverse V2 cores and an H100 GPU featuring 96 GB of HBM3 (3.4 TB/s) running Ubuntu 24.04.3 and CUDA 13.0.
    
    \item \textbf{System C}: Intel\textsuperscript{\textregistered} Xeon\textsuperscript{\textregistered} W9-3595X CPU featuring 60 cores and 256 GB of DDR5 (300 GB/s) running AlmaLinux 10.1.
\end{itemize}

The GPU-based Systems A and B represent distinct architectural trade-offs: while System B provides significantly higher memory bandwidth (HBM3 vs. GDDR7), System A features approximately 50\% more CUDA cores.
This disparity is important for the analysis, as it allows for the differentiation between compute-bound and memory-bound algorithms.
System C was specifically chosen for its superior multi-core CPU performance and serves as the dedicated test bed for evaluating the CPU-based reference implementations.

To provide a comprehensive analysis, Cuckoo-GPU is compared against the following data structures:

\begin{itemize}
    \item \textbf{GPU Blocked Bloom filter (GBBF)}: Sourced from the \textit{cuCollections} library \cite{cuCollections} which is based on WarpCore \cite{junger2020warpcore}, serving as a high-performance baseline for an append-only filter.

    \item \textbf{Partitioned CPU Cuckoo filter (PCF)}: A multi-threaded optimized variant \cite{schmidt2021four} of the original CPU Cuckoo filter \cite{cuckoo-filter} using 120 threads on System C.

    \item \textbf{Bulk Two-Choice filter (TCF)}: A modern, GPU-focused AMQ data structure that serves as a direct competitor \cite{gpu-filters}.

    \item \textbf{GPU Counting Quotient filter (GQF)}: A highly space-efficient probabilistic data structure sourced from the same study as the TCF \cite{gpu-filters}.

    \item \textbf{Bucketed Cuckoo Hash Table (BCHT)}: A high-performance, static GPU-based Cuckoo hash table sourced from \cite{awad2023analyzing}. 
    We include it as a baseline to demonstrate that while a full hash table can be used as a filter, it uses far more memory and yields lower throughput compared to a dedicated filter.
\end{itemize}

\subsection{Throughput Analysis}

To evaluate the architectural scalability of the filters, we test two distinct memory scenarios: an \textit{L2-resident} scenario, where the filter is small enough to fit entirely within the GPU's L2 cache ($2^{22}$ slots), and a \textit{DRAM-resident} scenario, where the filter's memory footprint significantly exceeds the cache capacity ($2^{28}$ slots), forcing accesses to global memory.
For our synthetic benchmarks, the input datasets consist of uniformly distributed 64-bit integers. All filters are configured to use 16-bit fingerprints, or an equivalent space allocation of 16 bits per item for the Blocked Bloom filter to ensure a fair comparison.
The only exception to this synthetic data generation is in Section \ref{sec:eval:kmer}, where we evaluate the filters using a real-world human genomic dataset.

\begin{figure*}
    \centering
    \includegraphics[width=0.75\linewidth]{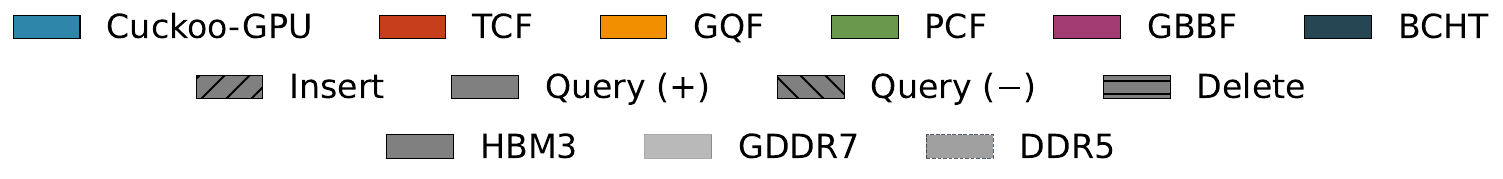}
    \begin{subfigure}[t]{0.49\linewidth}
        \centering
        \includegraphics[width=\linewidth]{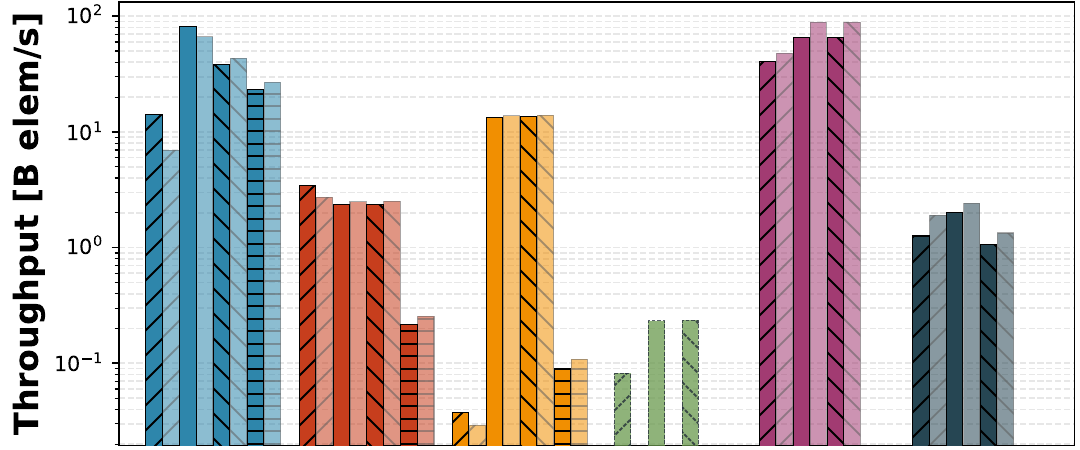}
        \caption{L2-resident filters of size 8 MiB.}
    \end{subfigure}
    \begin{subfigure}[t]{0.49\linewidth}
        \centering
        \includegraphics[width=\linewidth]{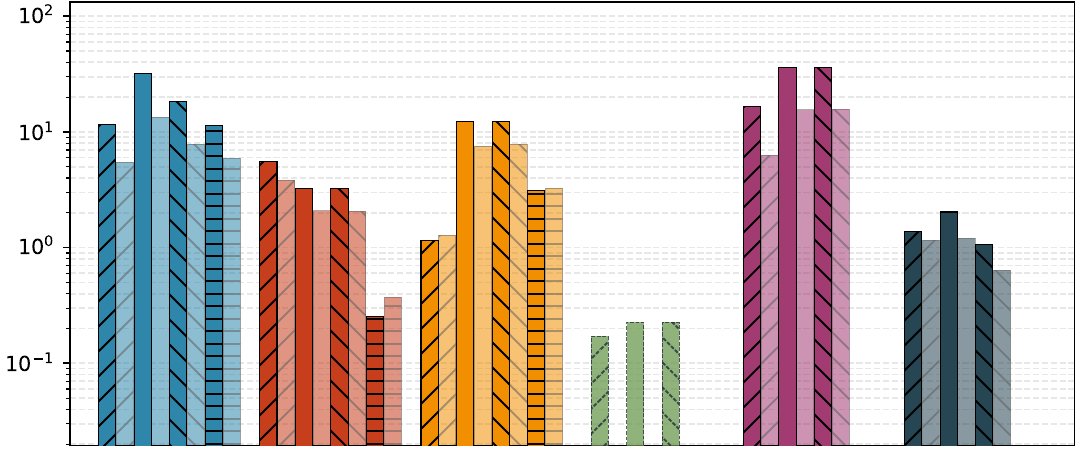}
        \caption{DRAM-resident filters of size 512 MiB.}
    \end{subfigure}
    \caption{Throughput for various filters and operations on Systems B (HBM3) and A (GDDR7) for all GPU-based implementations and System C (DDR5) for the CPU-based PCF at a constant 95\% target load factor.}
    \Description{This bar chart compares the throughput (in billions of elements per second, log scale) of six filter implementations: Cuckoo-GPU (blue), TCF (red), GQF (orange), PCF (green), BBF (purple), and BCHT (dark blue), across four operations: Insert, Query (+), Query (-), and Delete. The chart is split into two subplots:
    Left subplot (L2-resident): Cuckoo-GPU achieves the highest throughput for Insert (≈10 B elem/s) and Query (+) (≈100 B elem/s), outperforming all others but BBF by 1–2 orders of magnitude. TCF and GQF show moderate performance in Query (+) (≈5–10 B elem/s). PCF and BCHT are consistently low (≤2 B elem/s) across all operations.  
    Right subplot (DRAM-resident): BBF leads in all operations (≈20–40 B elem/s on HBM3), while Cuckoo-GPU is only slightly slower (≈10–30 B elem/s on HBM3). TCF and GQF maintain moderate performance (≈5–10 B elem/s). PCF and BCHT remain low (≤2 B elem/s).
    The chart illustrates that Cuckoo-GPU excels in high-bandwidth L2-resident scenarios, while BBF is slightly more competitive in DRAM-resident settings.}
    \label{fig:throughput-comparison}
\end{figure*}

Our primary performance metric is operation throughput, measured in billions of processed elements per second (B elem/s) for insertions, queries, and deletions.
To ensure statistical significance and mitigate system noise, every benchmark executes several warm-up iterations internally, and each benchmark configuration is independently run multiple times.
The results, presented Figure \ref{fig:throughput-comparison}, represent the overall median throughput across all runs for a constant 95\% target load factor.

\textbf{Insertion Performance.}
The insertion results highlight the severe cost of algorithmic complexity on GPU throughput.
In the L2-resident scenario, memory latency is largely hidden, exposing the compute and synchronisation bottlenecks of the dynamic filters.
Here, Cuckoo-GPU achieves a massive 378$\times$ speedup over the GQF and is 4.1$\times$ faster than TCF on System B (HBM3).

In the DRAM-resident scenario, the workload becomes bound by global memory bandwidth.
Cuckoo-GPU continues to dominate the dynamic alternatives, outperforming GQF by 10$\times$ (4.3$\times$) and TCF by 2.1$\times$ (1.4$\times$) on System B (System A).
The difference between memory architectures confirms that our Cuckoo filter does a much better job at utilising the massive HBM3 bandwidth, whereas TCF and GQF stagnate due to internal bottlenecks.
While Cuckoo-GPU trails GBBF (0.71$\times$ on System B, 0.87$\times$ on System A), this is expected as GBBF is an append-only structure that doesn't need to handle collisions.

\textbf{Query Performance.}
Query throughput demonstrates the efficiency of our bucketed layout and vectorised loads.
For positive queries (items present in the filter) in the L2-resident workload, Cuckoo-GPU even surpasses GBBF by 1.25$\times$ on System B.
This happens because the larger bucket size of 16 allows a positive query to often finish after a single memory transaction, whereas GBBF always has to consider the entire block.
Notably, negative queries on the Cuckoo filter exhibit a compute bottleneck, as having to traverse both buckets puts a much larger strain on the compute pipeline due to the bitwise arithmetic involved in our SWAR lookup.
Against the dynamic filters, we are 6$\times$ and 34.7$\times$ faster than GQF and TCF, respectively.

For DRAM-resident workloads, Cuckoo-GPU effectively matches GBBF for positive queries (0.90$\times$ on System B, 0.86$\times$ on System A) while maintaining a comfortable lead over GQF (2.6$\times$) and TCF (9.9$\times$).
Negative queries halve the throughput due to the memory bottleneck and needing to fetch both buckets, but Cuckoo-GPU remains significantly faster than all the other dynamic GPU filters.

\textbf{Deletion Performance.}
Because deleting an item merely requires a single atomic CAS, Cuckoo-GPU achieves a massive 258$\times$ speedup over GQF and 107$\times$ over TCF in L2-resident workloads on System B.
Even when bottlenecked by DRAM bandwidth, Cuckoo-GPU retains a 3.6$\times$ and 44.48$\times$ speedup over GQF and TCF, respectively.

\textbf{Hash Table and CPU Baseline.}
We also evaluated the viability of repurposing a full hash table for set membership.
BCHT was found to require approximately an order-of-magnitude more memory than the Cuckoo filter, significantly reducing the effective memory bandwidth utilisation.
As a result, Cuckoo-GPU outperforms it by 8.5$\times$ to 41$\times$ across all operations on System B, proving that a dedicated fingerprint-based filter is more efficient.

Finally, we compared GPU-Cuckoo against the multi-threaded PCF running on System C.
Across our tested architectures, Cuckoo-GPU delivers substantial performance gains over PCF, ranging from a 32.4$\times$ speedup for DRAM-resident insertions on System A, up to a 350$\times$ speedup for L2-resident positive queries on System B.

\subsection{False Positive Rate}

\begin{figure}
  \centering
  \includegraphics[width=\linewidth]{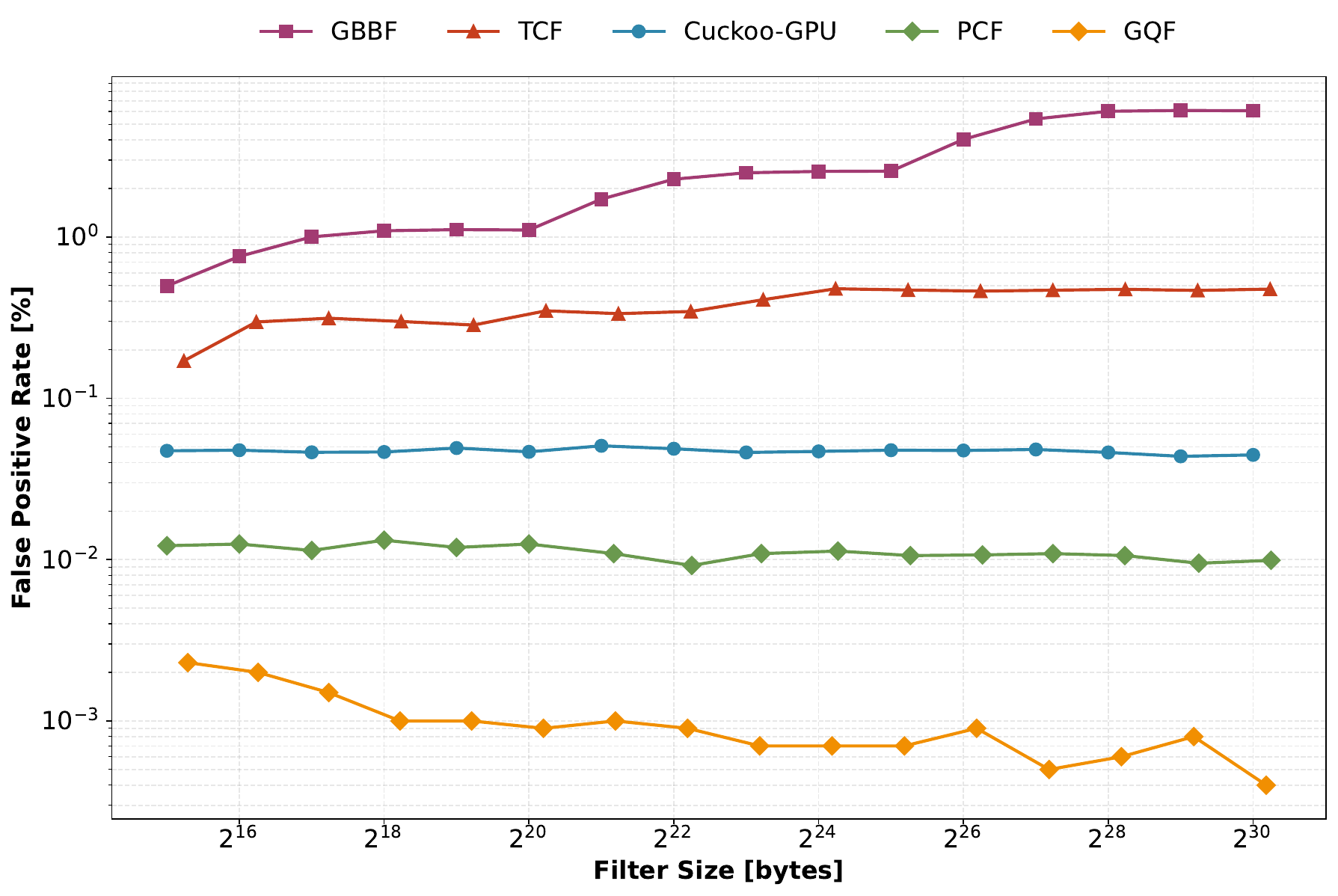}
  \caption{Comparison of False Positive Rates (FPR) versus total memory size for various filter implementations at a 95\% load factor. The L2-resident and DRAM-resident workloads evaluated in Figure \ref{fig:throughput-comparison} correspond to $2^{23}$ (8 MiB) and $2^{29}$ (512 MiB) bytes, respectively.}
  \Description{The chart shows how FPR changes as memory size increases for five filter types: BBF, TCF, Cuckoo-GPU, PCF, and GQF. Cuckoo-GPU keeps its FPR nearly constant at about 0.05\% across all memory sizes. PCF stays at around 0.01\%. GQF starts at about 0.0025\%, dips slightly to 0.001\% at 0.25 megabyte, then settles to 0.0004\% at 1 gigabyte. BBF begins at 0.5\% and steadily climbs to 1.6\% as memory grows. TCF starts at 0.15\% and rises gradually to 0.5\%.}
  \label{fig:fpr-vs-memory}
\end{figure}

To evaluate the reliability of the implemented filters, the empirical FPR was measured across a range of filter capacities.
This is done by first populating the filter with keys from the range $[0, 2^{32}-1]$ until a 95\% load factor is reached.
Subsequently, we query the filter using a distinct set of keys drawn from the disjoint range $[2^{32}, 2^{64}-1]$.
We calculate the empirical FPR as the fraction of these non-existent keys that incorrectly return a positive result.
The total memory size was varied from $2^{15}$ to $2^{30}$ bytes, allowing each implementation to optimise its internal layout within that fixed memory constraint.
To contextualise these results shown in Figure \ref{fig:fpr-vs-memory} with our throughput evaluation, the L2-resident ($2^{22}$ slots) and DRAM-resident ($2^{28}$ slots) configurations correspond to approximately 8 MiB ($2^{23}$ bytes) and 512 MiB ($2^{29}$ bytes) on the x-axis, respectively.

\begin{itemize}
    \item \textbf{Blocked Bloom filter}: The Blocked Bloom filter demonstrates the highest FPR among all tested structures, ranging from approximately 0.5\% to 6\%.
    This is a known characteristic of the blocked design: partitioning the bit array into small, fixed-size blocks prevents the "averaging" of hash collisions across the entire filter.
    Consequently, a few heavily congested blocks can disproportionately skew the overall error rate.
    It is notably the only filter where the FPR degrades this much as the total memory size increases.

    \item \textbf{Quotient filter}: The GQF exhibits the lowest FPR among all candidates, maintaining an error rate below 0.002\%.
    This confirms the theoretical space efficiency of quotient filters, which handle collisions via Robin Hood hashing and metadata encoding.

    \item \textbf{GPU vs. CPU Cuckoo filters}: A distinction remains visible between the CPU and GPU Cuckoo filter implementations.
    The CPU version achieves a very low FPR, hovering near 0.005\%.
    The GPU Cuckoo filter, while still highly accurate, has a higher rate of approximately 0.045\%.
    To maximise throughput, the GPU implementation uses a bucket size of 16, whereas the CPU versions use a standard bucket size of 4.
    As established in Equation \ref{eq:cuckoo-fpr}, a larger bucket size directly increases the collision probability for a fixed fingerprint size.
    It is important to note that, unlike the CPU versions, the GPU Cuckoo filter allows for flexible bucket sizes to tune the FPR.

  \item \textbf{Comparison with TCF}: The GPU Cuckoo filter significantly outperforms the TCF regarding accuracy.
    The TCF has an error rate roughly an order of magnitude higher (ranging between 0.35\% and 0.55\%).
    While the TCF is more accurate than the Blocked Bloom filter, the Cuckoo filter and GQF designs offer superior accuracy for this workload.
\end{itemize}

\subsection{Algorithmic Optimizations}

\subsubsection{\textbf{Eviction Policies}}
\label{sec:eval:eviction-policies}

To evaluate the effectiveness of our BFS eviction heuristic, we compare it against the standard greedy DFS approach. 
We focus our analysis on the DRAM-resident scenario ($2^{28}$ slots) on the GH200. 
(Note that for L2-resident filters, memory latency is low enough that the choice of eviction policy has no meaningful impact on overall throughput). 
To accurately capture performance under heavy contention, we isolate the measurement to the final phase of construction.
Specifically, to reach a target load factor $\alpha$, we first pre-fill the filter with three-quarters of the required items.
We then only measure the throughput of inserting the final quarter of the items.

\begin{figure}
    \centering
    \includegraphics[width=\linewidth]{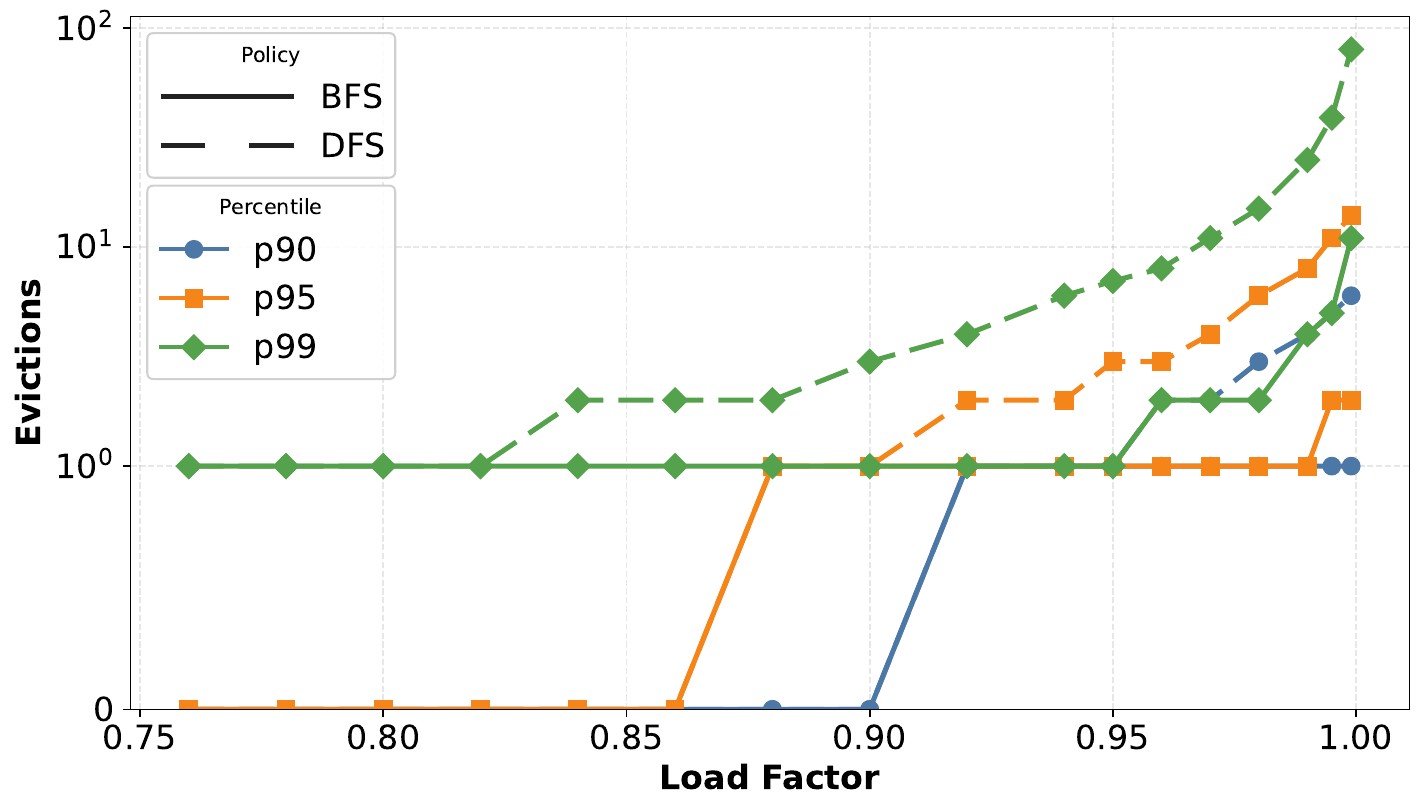}
    \caption{Tail eviction counts (90th, 95th and 99th percentiles) per insertion for BFS vs. DFS on System B. }
    \Description{The line chart illustrates the tail eviction counts for the 90th, 95th, and 99th percentiles per insertion for Breadth-First Search (BFS) and Depth-First Search (DFS) on System B, as a function of load factor ranging from 0.75 to 1.0. The y-axis, presented on a logarithmic scale, quantifies the number of evictions, while the x-axis represents the load factor.

For BFS, eviction counts remain at 1 for the 99th percentile until a load factor of approximately 0.95, after which they increase up to a plateau at roughly 2 between a load factor of 9.6 and 9.8, then increases sharply to 10 at a load factor of 1. The 95th percentile starts at 0 evictions and begins to rise at a load factor of approximately 0.87, also increasing to 10 at a load factor of 1, while the 90th percentile remains at 0 until a load factor of approximately 0.90, after which it increases to 1.

For DFS, the 99th percentile remains at 1.0 until a load factor of approximately 0.82, plateaus at 2 between 8.4 and 8.7 after which it rises steeply to 100 at a load factor of 1. The 95th percentile begins at 0 and starts to increase at a load factor of approximately 0.87 and reaches 10 at a load factor of 1, while the 90th percentile remains at 0 until a load factor of approximately 0.90, after which it also increases and reaches approximately 5 at a load factor of 1.}
    \label{fig:evict_percentiles}
\end{figure}

Figure \ref{fig:evict_percentiles} compares the 90th, 95th, and 99th percentiles of eviction chain lengths per insertion for both policies. 
While both perform similarly at lower load factors, the DFS strategy experiences an exponential explosion in tail evictions as the filter approaches capacity. 
Conversely, the BFS strategy drastically suppresses these outliers. 
By exhaustively searching for shallow eviction paths before extending the depth of the chain, BFS effectively bounds the worst-case sequential memory accesses.

This reduction in eviction depth presents a direct trade-off: BFS performs more total memory reads (to check multiple candidates) but significantly reduces serial pointer-chasing and atomic writes. 

\begin{figure}
    \centering
    \includegraphics[width=\linewidth]{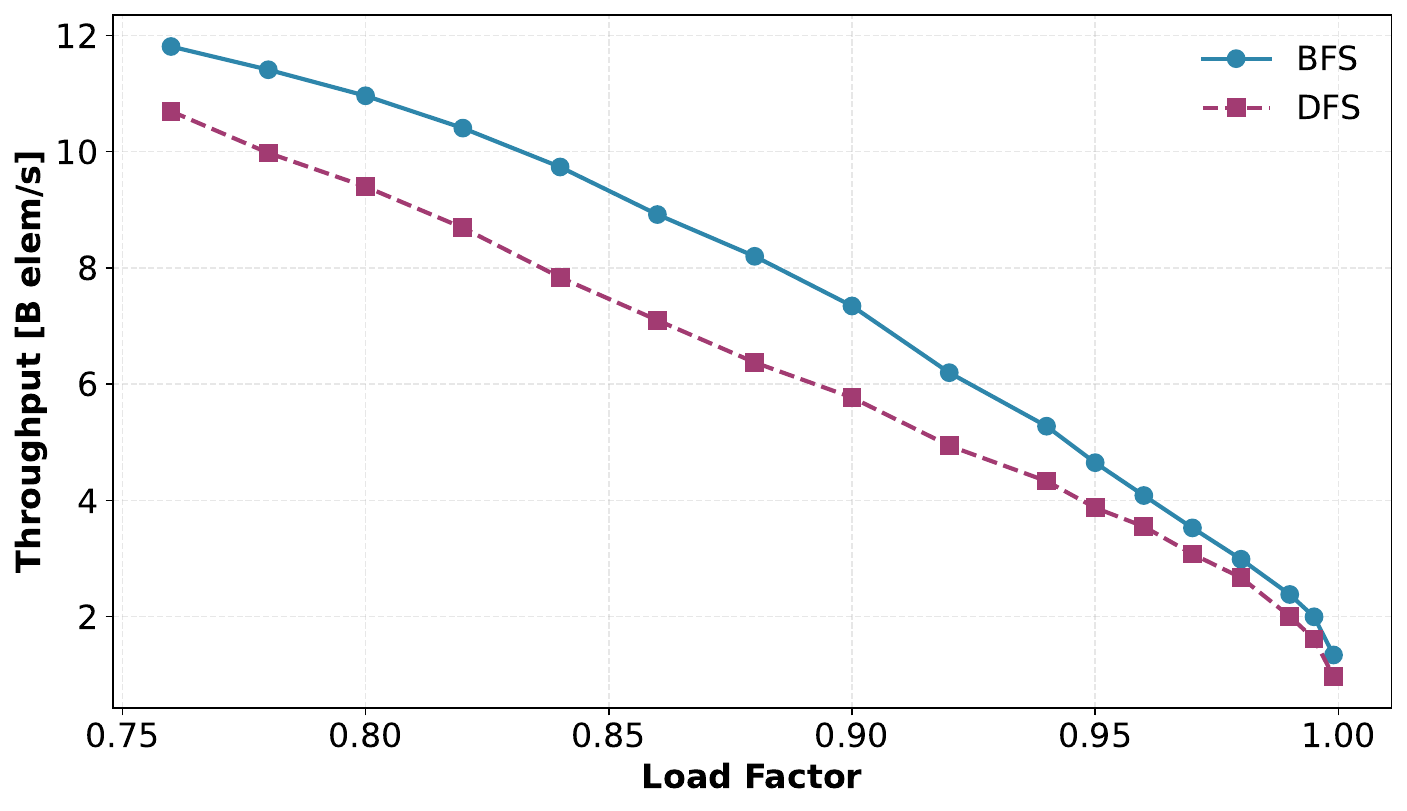}
    \caption{Insertion throughput comparison of BFS and DFS eviction policies on System B (DRAM-resident).}
    \label{fig:evict_throughput}
    \Description{The line chart presents a comparative analysis of insertion throughput (measured in billion elements per second). The x-axis represents the load factor, ranging from 0.75 to 1.0, while the y-axis quantifies throughput in billion elements per second between 0 and 12.

Breadth-First Search (BFS) demonstrates consistently higher insertion throughput across all load factors compared to Depth-First Search (DFS). At a load factor of 0.75, BFS achieves approximately 12 billion elements per second, while DFS achieves approximately 11 billion elements per second. As the load factor increases, both policies exhibit a monotonic decline in throughput, with BFS maintaining a performance advantage of approximately 1–2 billion elements per second throughout most of the observed range.

At a load factor of 1, BFS throughput drops to approximately 1.5 billion elements per second, while DFS throughput falls to approximately 1.0 billion elements per second.}
\end{figure}

Figure \ref{fig:evict_throughput} illustrates the impact on insertion throughput. 
On the GH200, the baseline DFS policy becomes severely bottlenecked by latency during long eviction chains, as threads stall waiting for sequential global memory accesses to resolve. 
The BFS policy breaks this dependency. It replaced deep, serial chains of atomic updates with a series of simple probes to candidate buckets. 
Because modern GPUs like the GH200 provide massive memory bandwidth but remain highly sensitive to latency stalls, they can absorb the extra read requests efficiently while penalising dependent atomics. 
By trading this abundant memory bandwidth for a drastic reduction in latency-bound atomic operations, the BFS approach maintains significantly higher and more stable performance as the filter fills up, outperforming DFS by up to 25\%.

\subsubsection{\textbf{Bucket Policies}}
\label{sec:eval:bucket-policies}

\begin{figure}
    \centering
    \includegraphics[width=\linewidth]{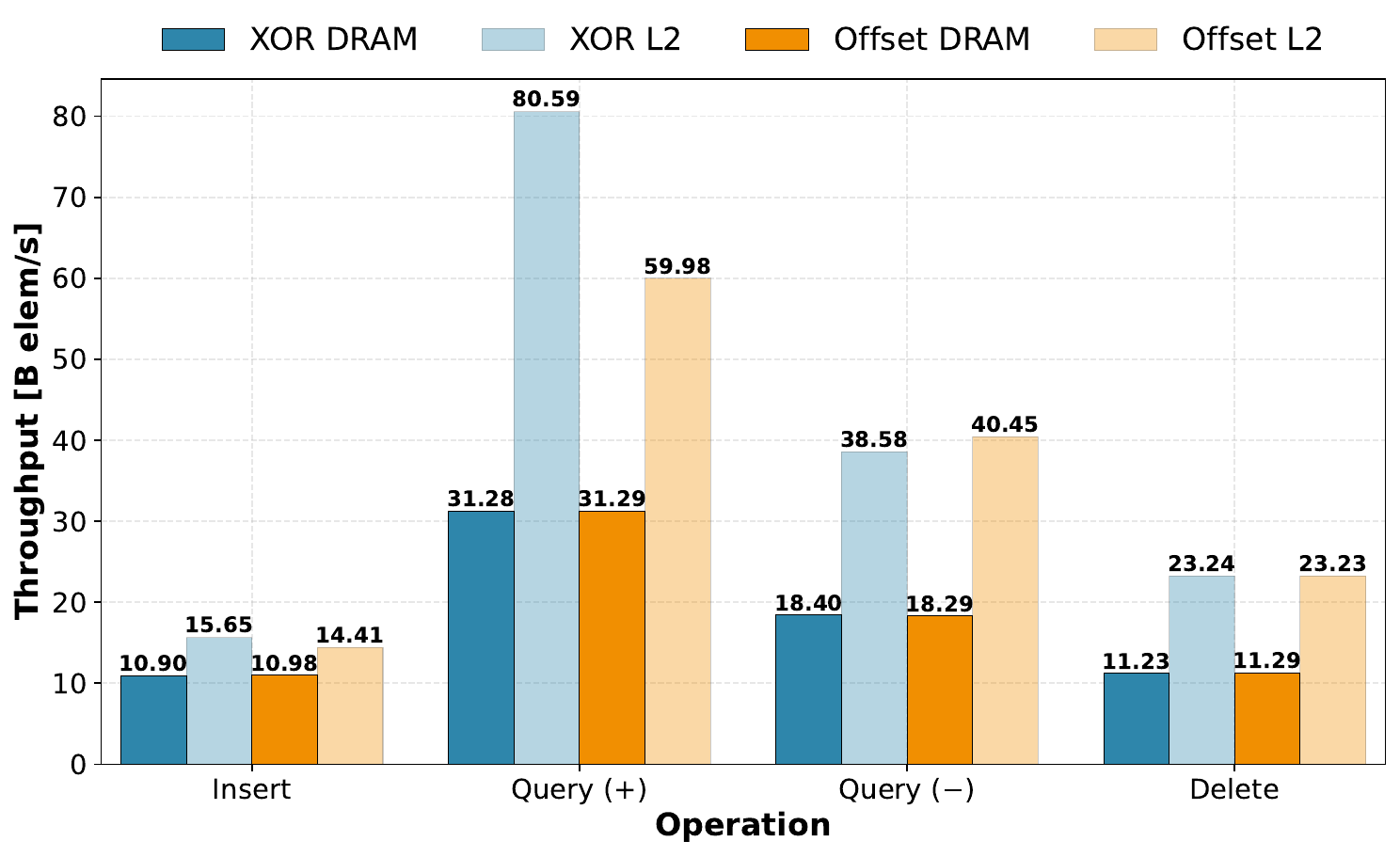}
    \caption{Performance of the various Bucket Policies on System B}
    \label{fig:bucket-policies}
    \Description{The bar chart compares throughput in billion elements per second of the two bucket policies XOR and Offset on L2-cache- and DRAM-resident versions on System B across Insert, Query (+), Query (-), and Delete operations.
    For Insert: XOR L2 achieves 15.65, XOR DRAM 10.90, Offset L2 14.41, and Offset DRAM 10.98..
    For Query (+): XOR L2 peaks at 80.59, followed by Offset L2 at 59.98, XOR DRAM at 31.28, and Offset DRAM at 31.29.
    For Query (-): Offset L2 reaches 40.45, XOR L2 38.58, XOR DRAM 18.40, and Offset DRAM 18.29.
    For Delete: XOR L2 and Offset L2 are nearly tied at 23.24 and 23.23, respectively, while XOR DRAM and Offset DRAM trail at 11.23 and 11.29.
    L2-resident policies consistently deliver higher throughput, with XOR L2 dominating Query (+) and both L2 policies performing similarly in Delete.}
\end{figure}

As detailed in Section \ref{sec:impl:bucket-policies}, the standard XOR-based partial-key Cuckoo hashing imposes a strict power-of-two constraint on the number of buckets.
To evaluate the cost of flexibility, the standard XOR policy was benchmarked against the Offset (Choice-bit) policy on System B with a fixed load factor of 95\%.
The results for both L2-resident and DRAM-resident scenarios are presented in Figure \ref{fig:bucket-policies}.

The performance trade-off between the standard XOR policy and the flexible Offset-based policy depends strictly on memory residency.
In L2-resident scenarios, the filter is bound by instruction latency; here, the XOR policy's use of simple bitwise masking makes it approximately 34\% faster for positive queries than the Offset policy, which requires more expensive modulo arithmetic.
Because VRAM capacity is rarely a constraint for small filters, XOR remains the optimal choice.

Conversely, in DRAM-resident scenarios, the workload becomes completely memory-bound.
The computational overhead of the Offset calculation is entirely hidden by global memory latency, allowing it to perfectly match the throughput of the XOR baseline across all operations.

Consequently, for large datasets, the Offset policy is a highly compelling alternative.
By eliminating the strict power-of-two sizing restriction, it can prevent massive memory over-provisioning in exchange for a single bit of fingerprint entropy.

\subsection{Case Study}
\label{sec:eval:kmer}

While synthetic benchmarks using uniformly distributed integers are useful for showing algorithmic behaviour, real-world data can often present challenges due to skewed distributions.
To validate this, a benchmark was conducted on genomic $k$-mer indexing as an important use case for approximate membership query structures in bioinformatics.

A $k$-mer is a substring of length $k$ derived from a biological sequence, such as DNA.
In genomic analysis, counting and filtering $k$-mers is an important step for tasks like genome assembly and error correction.
Since DNA consists of four bases (A, C, G, T), the total number of possible $k$-mers is $4^k$, which grows rather quickly.

For this evaluation, the T2T-CHM13 
dataset is used as the first complete sequence of a human genome.
The raw FASTA data was pre-processed using KMC3 \cite{kmc3} to extract all distinct 31-mers.
To optimise memory usage and processing speed, the text-based $k$-mers were packed into a 2-bit-per-base binary representation, compressing the dataset roughly by a factor of four and allowing each 31-mer to fit within a single \texttt{uint64\_t}.

The resulting dataset is approximately 20 GB in size (packed), which is just enough to fully saturate the 96 GB of VRAM on System B while benchmarking.
All filters were tested for insertion, positive queries, and deletion (when supported).

\begin{figure}
  \centering
  \includegraphics[width=\linewidth]{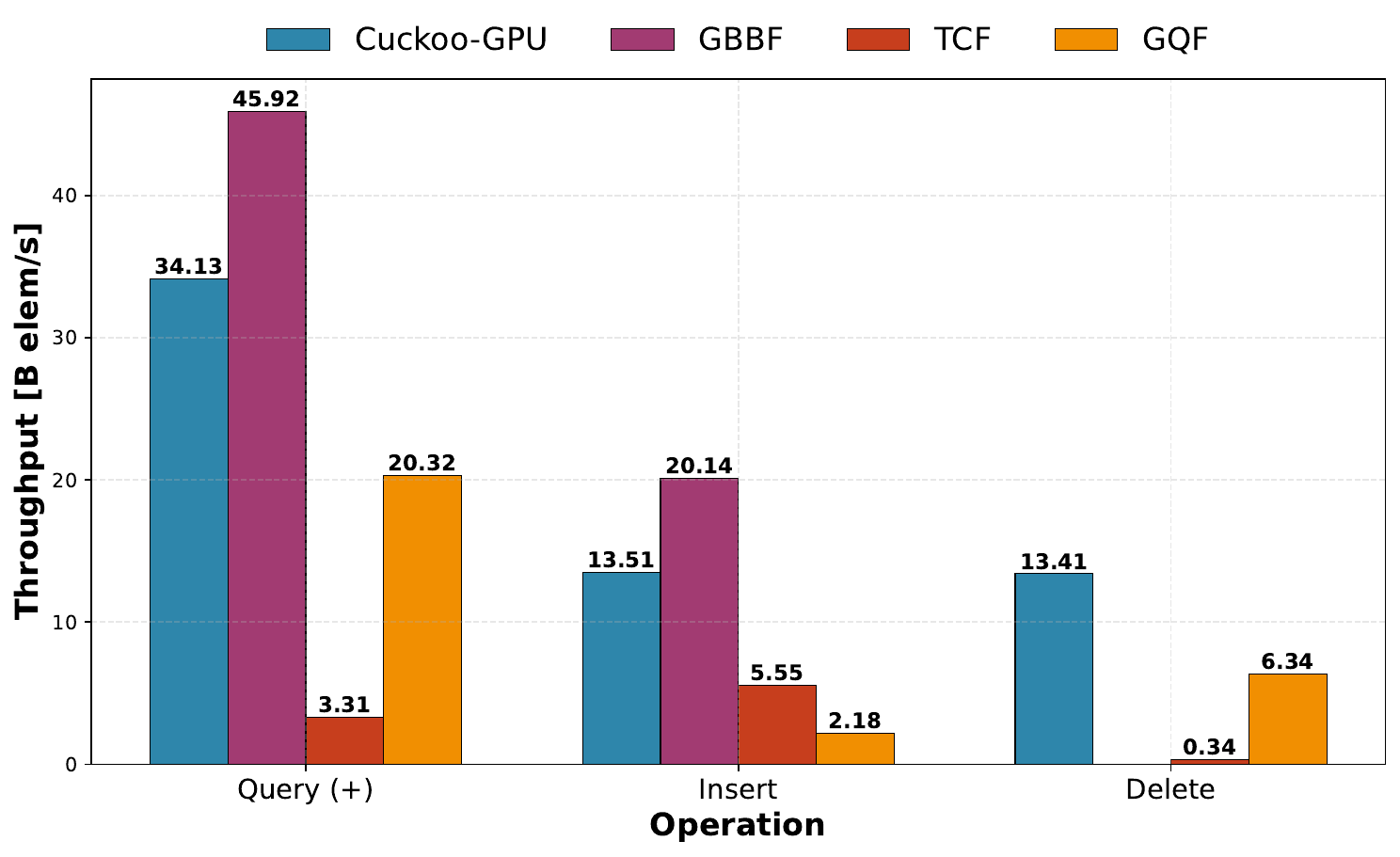}
  \caption{Throughput comparison for inserting, querying, and deleting 31-mers from the full T2T-CHM13 human genome on System B.}
  \Description{The bar chart compares throughput in billion elements per second across four data structures, Cuckoo-GPU, GBBF, TCF, GQF, for Query (+), Insert, and Delete operations.
  Query (+): GBBF achieves 45.92 B elem/s, the highest among all structures. Cuckoo-GPU follows at 34.13, GQF at 20.32, and TCF at 3.31.
  Insert: GBBF leads at 20.14 B elem/s, followed by Cuckoo-GPU at 13.51, TCF at 5.55, and GQF at 2.18.
  Delete: Cuckoo-GPU leads at 13.41 B elem/s, followed by GQF at 6.34, GBBF at 0.34, and TCF has no delete operation.}
  \label{fig:kmer-benchmark}
\end{figure}

The throughput results for the $k$-mer benchmark are presented in Figure \ref{fig:kmer-benchmark}.
They confirm that the Cuckoo filter's high performance translates well to real-world workloads:

\begin{itemize}
    \item \textbf{Insertion}: While the Cuckoo filter trails the append-only Blocked Bloom filter in this high-bandwidth scenario, it remains the fastest among dynamic data structures, outperforming the TCF by 2.4$\times$ and the GQF by 6.2$\times$.
  
    \item \textbf{Query}: While the Cuckoo filter does not fully match the Blocked Bloom filter, it maintains a notable lead over other dynamic alternatives.
    It is 68\% faster than the GQF and 10.3$\times$ faster than the TCF.

    \item \textbf{Deletion}: In deletion throughput, the Cuckoo filter demonstrates superior performance.
    It is 2.1$\times$ faster than the GQF and 39.2$\times$ faster than the TCF.
\end{itemize}

\section{Conclusion}
\label{sec:conclusion}

We have presented {\it Cuckoo‑GPU}, an open‑source Cuckoo‑filter library available at \url{\repourl} that is tuned for modern GPU architectures. Earlier GPU‑based AMQ designs attempted to eliminate the random‑access pattern of Cuckoo hashing by introducing complex sorting or shifting steps; these techniques, however, incurred high latency and synchronization overheads. By contrast, Cuckoo‑GPU embraces Cuckoo hashing’s inherent randomness, using a lock‑free, atomic CAS scheme that fully saturates the global memory bandwidth of devices such as the NVIDIA GH200 (HBM3). Our novel BFS eviction heuristic shortens eviction chains, keeping throughput high even at elevated load factors.

Comprehensive experiments show that Cuckoo‑GPU closes the performance gap between static (append‑only) and dynamic filters. Insertion and deletion throughputs exceed the current state-of-the-art by one-to-two orders-of-magnitude, while query performance rivals the highly optimized Blocked Bloom filter. The gains observed in a genomic $k$-mer indexing case study translate directly into substantial speedups for real‑world workloads.

Despite these advances, Cuckoo filters retain intrinsic limitations. At extreme load factors the data structure can experience insertion failures, which necessitates fallback mechanisms. 
Moreover, tuning the false positive rate is constrained to coarse, discrete steps.
Because the FPR is dictated by both the fingerprint size and the bucket capacity --- which our implementation restricts to hardware-friendly widths (e.g., 8, 16, or 32 bits) and strict powers of two --- users cannot finely control it like they can with a Bloom filter.
Addressing these edge cases and constraints through more resilient insertion strategies and false‑positive tuning is an exciting avenue for future work.

In summary, Cuckoo‑GPU shows that dynamic updates can now be supported without major throughput penalties, thereby unlocking new scaling opportunities for next‑generation data processing pipelines on modern accelerators.

\printbibliography
\end{document}